# REPRESENTATION FOR ALPHANUMERIC DATA TYPE BASED ON SPACE AND SPEED CASE STUDY: STUDENT ID OF X UNIVERSITY

Agus Pratondo

Department of Information Technology, Telkom Polytechnic, Bandung, Indonesia
agus@politel.ac.id

## ABSTRACT

*ID is derived from the word identity, derived from the first two characters in the word. ID is used to distinguish between an entity to another entity. Student ID (SID) is the key differentiator between a student with other students. On the concept of database, the differentiator is unique. SID can be numbers, letters, or a combination of both (alphanumeric). Viewed from the daily context, it is not important to determine which a SID belongs to the type of data. However, when reviewed on database design, determining the type of data, including SID in this case, is important. Problems arise because there is a contradiction between the data type viewed from the data characteristic and practical needs. Type of data for SID is a string, if it is evaluated from the basic concepts and its characteristic. It is acceptable because SID consists of a set of numbers which will not be meaningful if applied arithmetic operations like addition, subtraction, multiplication and division. But in terms of computer organization, data representation type will determine how much data space requirements, speed of access, and speed of operation. By considering the constraints of space and speed on the experiments conducted, SID is better expressed as an integer rather than a set of characters.*

## 1. INTRODUCTION

X University currently has a number of students about 3600 students. As a higher education that implements Information and Communication Technology (ICT), X University has an Academic Information System to manage all activities related to academic, good facilities and infrastructure. All data relating to the academic process is stored in a Database Management System that integrates with various applications.

Student is an entity which has a large amount totally and continue to grow from time to time. Various attributes of the entity are attached to the student and the total are 46 attributes. Table 1 below describes some attributes attached to a student.

Table 1.  The Atributes of A Student

| No | Field | Type |
|----|-------|------|
| 1 | SID | varchar(10) |
| 2 | UNIT_CODE | varchar(100) |
| 3 | NAME | varchar(255) |
| 4 | PLACE_BIRTH | varchar(100) |
| 5 | DATE_BIRTH | Date |
| 6 | SEX | char(1) |
| ... | ... | ... |
| 46 | PRIM_CLASS | varchar(100) |

In addition, the amount of data students continue to increase each year as new admissions. The following table shows the growth significantly from year to year since 2007.

Table 2.  Student Body of X University

| Year | Department | | | Total |
|------|-----------|-----------|-------------|-------|
| | Comp. Acc. | Comp. Eng. | Inf. System | |
| 2007 | 65 | 150 | 207 | 422 |
| 2008 | 157 | 380 | 593 | 1130 |
| 2009 | 122 | 337 | 486 | 945 |
| 2010 | 121 | 328 | 422 | 871 |

Academic Information System of X University can serve a variety of daily academic process. However, several access for certain web pages are still slow and need to be improved for better performance.

An important issue in the improvement of the Academic Information System is Database design improvement. In the existing database structure, Student ID (SID)  which is the primary key for the entity of student and became an important key in table operation of database manipulation, is still represented as strings. SID is a set of alphanumeric digits that indicate the personal identity of a student. SID contains of information of the level degree, the study program, the entry year, and the serial number for a certain year.

Alphanumeric is a series of numbers but it does not have semantics value as a number. Thus, alphanumeric data should not be given arithmetic operators such as addition, subtraction, multiplication, and division. For example, the addition operation SID, SID multiplication with a certain number, or find the average of SID at a particular class will not give any meaning.

This paper will discuss which data type should be chosen for alphanumeric by using SID as case study. The considerations are based on  performance speed of the process and the use of storage space.

## 2. DATA REPRESENTATION

All data on the computer basically expressed as binary numbers which are called bits (derived from the word binary digits), such as number in base 2 consisting of the digits 0 and 1. Bit representation is the simplest representation in number systems when implemented in electronic circuits. The difference values 0 and 1 can be identified by the presence or absence of electric current with a certain threshold limit [5-12].

In the representation of data, units used for a number of bits of data are bytes, where 1 byte equals to 8 bits of data. All digital data on a computer are represented by bytes of data. The more bytes of data are used, the larger the data can be represented. For example, there are n-bit vector as follows:

$$B = b_{n-1}b_{n-2} \ldots b_1 b_0$$

where $b_i = 0$ or 1 for $0 \le i \le n-1$, then this vector can represent unsigned integer V in the range 0 to $2^n - 1$. The value of V in base 10 can be calculated by applying the following formula:

$$V(B) = \sum_{i=1}^{n} b_{n-i} \times 2^{n-i} \qquad (1)$$

The description of 1 byte data can be viewed at the following figure:

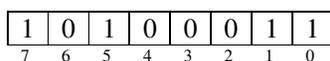

Figure 1. A Byte of Bits

In the figure 1, an integer that is represented by 1 byte data with a value 10100011 in base 2 (binary) can be calculated the value in base 10 (decimal) as follows:

$$
\begin{aligned}
10100011_2 \quad &= 1\text{x}2^7 + 0\text{x}2^6 + 1\text{x}2^5 + 0\text{x}2^4 \\
&\quad + 0\text{x}2^3 + 0\text{x}2^2 + 1\text{x}2^1 + 1\text{x}2^0 \\
&= 128 + 0 + 32 + 0 + 0 + 0 + 2 + 1 \\
&= 163_{10}
\end{aligned}
$$

The longer the byte representation of the data, the greater the value that will represent, but the greater the storage media will be required.

The following table descibes the representation of integers in My SQL[2].

Table 3. Representation of Integer in MySQL

| Type Data | Byte |
|-----------|------|
| TINYINT | 1 |
| SMALLINT | 2 |
| MEDIUMINT | 3 |
| INT | 4 |
| BIGINT | 8 |

Each integer data type consists of two types of numbers which are represented, ie Signed Integer (S / I) and Unsigned Integer (U / I). Each affects the value that represents as follows (Min = Minimum, Max = Maximum):

Table 4.  Sign Integer in MySQL

| No | Integer Type | Signed Integer (S/I) | |
|---|---|---|---|
| | | Minimum | Maximum |
| 1 | TINYINT | -128 | 127 |
| 2 | SMALLINT | -32768 | 32767 |
| 3 | MEDIUMINT | -8388608 | 8388607 |
| 4 | INT | -2147483648 | 2147483647 |
| 5 | BIGINT | -9223372036854775808 | 9223372036854775807 |

Table 5.  Unsign Integer in MySQL

| No | Integer Type | Unsigned Integer (U/I) | |
|---|---|---|---|
| | | Minimum | Maximum |
| 1 | TINYINT | 0 | 255 |
| 2 | SMALLINT | 0 | 65535 |
| 3 | MEDIUMINT | 0 | 16777215 |
| 4 | INT | 0 | 4294967295 |
| 5 | BIGINT | 0 | 18446744073709551615 |

MySQL recognizes not only integer data type, but also a string data type that is implemented as a CHAR and VARCHAR[1]. Both are used to store strings. The difference is how to represent the data. VARCHAR has additional attributes that will store the string length, while CHAR does not have this attribute. The length of the string that is stored by CHAR is always a fixed length. CHAR and VARCHAR types are declared by including a maximum length of characters that can be saved. For example, VARCHAR (30) means that total characters stored is maximum 30. The fixed length of CHAR  is defined once when the table strucutre is created. CHAR length is in the range 0 to 255. While VARCHAR length varies according to length of its stored data. The length of VARCHAR is in the range 0 to 255 for MySQL 5.0.3 and 0 up to 65,535 for the next version. In principle, VARCHAR stores data string plus one or two bytes. The bytes are used to save the information of string length. If the length of string is less than 255, VARCHAR will require only 1 byte, but if the characters are more than or equal to 255, it will require two bytes to store the string length. The following table describes the differences between the CHAR (4) and VARCHAR (4). Because the length is less then 255, VARCHAR requires a single byte addition in its representation.

Table 6.  The Data Representation of CHAR(4)

| Data | CHAR(4) | Jumlah Byte |
|---|---|---|
| `` `` `` | `` `    ` `` | 4 |
| `` `ab` `` | `` `ab  ` `` | 4 |
| `` `abcd` `` | `` `abcd` `` | 4 |
| `` `abcdefgh` `` | `` `abcd` `` | 4 |

Table  7.  The Data Representation of VARCHAR(4)

| Data | VARCHAR(4) | Jumlah Byte |
|---|---|---|
| `` `` `` | `` `    ` `` | 1 |
| `` `ab` `` | `` `ab  ` `` | 3 |
| `` `abcd` `` | `` `abcd` `` | 5 |
| `` `abcdefgh` `` | `` `abcd` `` | 5 |

## 2. THE DATA REPRESENTATION OF STUDENT ID (SID)

Alphanumeric representation is interesting to be explored because it may effect to access speed and disk space usage. In case of Student ID (SID), alphanumeric of SID  has some rule as follows:

1. It contains 8 digit.
2. The first digit represents the level degree, as follows:
   3 : for  Diploma 3
   4 : for  Diploma 4
3. The second and third digits represent the study program, as follows:
   01 : Information System
   02 : Computer Engineering
   03 : Computerization Accountancy
4. The fourth and fifth digits represent the entry year, as follows:
   08 : Year 2008
   09 : Year 2009
   10 : Year 2010
5. The sixth, seventh, and eighth digit represent the sequence in a certian  year.
   001 : number 1
   002 : number 2
   189 : number 189

SID, basically is an alphanumeric. In the concept of data, it should be expressed as a set of characters, which is represented as a VARCHAR in MySQL. In Academic Information System of X University, SID is represented by VARCHAR(10).

## 3. COMPARISON OF STORAGE SPACE

SID can be represented into alphanumeric data in MySQL as CHAR(8) or VARCHAR(8) as a sequence of characters. If this alphanumeric is represented as integers, it can be stored by considering the range of integer values, according to the integer type. Integer values to represent SID is the largest integer that can be built with 8 digits.

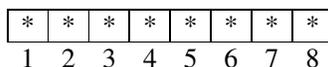

| * | * | * | * | * | * | * | * |
|---|---|---|---|---|---|---|---|
| 1 | 2 | 3 | 4 | 5 | 6 | 7 | 8 |

Figure 2. A Sequence of Number

The largest integer that can be formed from 8 single digits occurs if each single number for each * has the maximum value. The maximum value of one single digit is 9, so the maximum value of the integer representation for the SID is 99,999,999.

By considering the range limits of each integer, the available integer representation is described as follows:

Table 8.  The Representation  of 99.999.999 in MySQL

| Data Type | In the Range | Required Bytes |
|-----------|:------------:|:--------------:|
| TINYINT | No | 1 |
| SMALLINT | No | 2 |
| MEDIUMINT | No | 3 |
| INT | Yes | 4 |
| BIGINT | Yes | 8 |

By considering the available integers,CHAR, and VARCHAR, the required bytes can be compared as afollows:

Table 9.  SID Representation

| Data | Data Type | Required Bytes |
|------|-----------|----------------|
| `99999999` | CHAR(8) | 8 |
| `99999999` | VARCHAR(8) | 9 |
| 99999999 | INT | 4 |
| 99999999 | BIGINT | 8 |

From table 9, it can be infered that the minimum required bytes for SID are 4 bytes. It occurs when the data type is INT.  This type is the most efficient compared with other type. The efficiency for SID which is represented by INT reaches 50% compared to CHAR(8), and 56% compared to VARCHAR(8). Based on the case study in table 1, the efficiency of INT reachs 64% compared to VARCHAR(10).

## 4. COMPARISON OF SPEED ACCESS

Viewed from the use of storage space, data representation of SID by using integer is more efficient. However, another important dimension of quality in web applications is the speed of access. To simulate this access speed, two tables are made to compare the speed between SID as VARCHAR (8) and as INT. The structure of the tables is describe as follows:

Table 10.  Table SID_STR

| Field | Type |
|-------|------|
| sid | VARCHAR(8) |
| name | VARCHAR(100) |

Table 11.  Tabel SID_INT

| Field | Type |
|-------|------|
| sid | INT |
| name | VARCHAR(100) |

Each table is filled with the same data. In this experiment, the two tables filled with data as much as 100,000 with SID sorted from 10100000 to 10200000. Field "name" is filled with a uniform name for all data. To generate the data for the experiment, the following PHP script is applied[3].

```
$sid = 10100001;
while ($sid < 10200001){

$querystring = "insert into sid_str values ('".$sid."','Ananda
Putera Perkasa')";
$queryint   = "insert into sid_int values (".$sid.",'Ananda
Putera Perkasa')";
  $sid++;
}
```

Figure 2.Code to Generate Data

Furthermore, Each tables are filled with 100.000 data, then a SELECT query command is applied to access to the table with the SID with type VARCHAR (8) and INT. To get data as accurate as possible, the total SELECT query in each experiment is set up as much as possible. In this case,the total query is 100.000 commands [4].

```
function getmicrotime(){
list($usec, $sec) = explode(" ",microtime());
return ((float)$usec + (float)$sec);
}

$i=0;
$t1 =0;
$t2 =0;

while ($i < 100000){
$q="select * from sid_str where sid='10200000'";
$time_start = getmicrotime();
mysql_query($q);
$time_end = getmicrotime();
$time = $time_end - $time_start;
$t1=$t1+$time;
//INTEGER
$q="select * from sid_int where sid=10200000" ;
$time_start = getmicrotime();
mysql_query($q);
$time_end = getmicrotime();
$time = $time_end - $time_start;
$t2=$t2+$time;
$i++;
}
echo "TOTAL TIME FOR STRING: $t1";
echo "TOTAL FOR INTEGER    : $t2";
```

Figure 3.Code to Query The Tables

Finally, the result of the experiment is listed below:

Table 12.  Test Result

| No. | SID Type | |
|---|---|---|
| | VARCHAR(8) | INT |
| 1 | 23,735124826431 | 23,480888843536 |
| 2 | 24,585271120071 | 24,313096523285 |
| 3 | 23,454295873642 | 23,258988857269 |
| 4 | 24,026872873306 | 23,827021121979 |
| 5 | 23,368565320969 | 23,145508527756 |
| 6 | 23,883385658264 | 23,567368268967 |
| 7 | 23,488732099533 | 23,276761293411 |
| 8 | 24,059506416321 | 23,790077209473 |

| 9 | 23,642956495285 | 23,390990972519 |
| 10 | 23,850153207779 | 23,539158582687 |
| 11 | 23,538511991501 | 23,300233840942 |
| 12 | 23,894111633301 | 23,704314231873 |
| 13 | 23,469617605209 | 23,224228143692 |
| 14 | 23,740539550781 | 23,503800630569 |
| 15 | 24,282388448715 | 23,967983245850 |

The chart for the table above is described as follows:

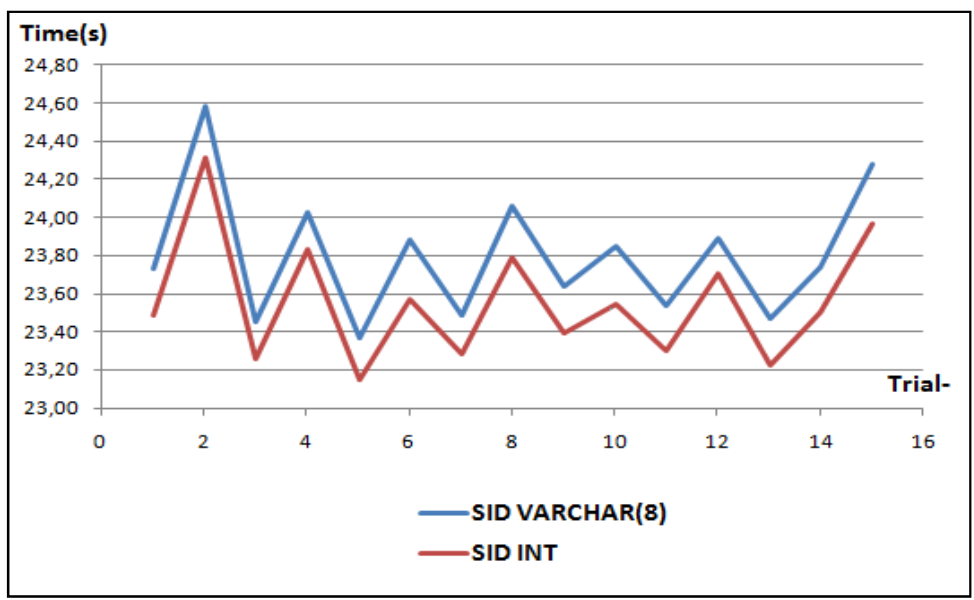

Figure 4.Query Time Results

The graph in Figure 4 shows that the total time for 100.000 query commands on the SID with type VARCHAR (8) located above the SID with the type INT. The meaning of this graph is the access time for the SID with type VARCHAR(8) is longer than that with the type INT.

The total time is 357,020033121108 for VARCHAR(8) and 353,290420293808 for INT. The efficiency of time using INT data type can be calculated as follows[9]:

$$\eta \quad = \frac{t_{VARCHAR(8)} - t_{INT}}{t_{VARCHAR(8)}}$$

$$= \frac{357,020033121108 - 353,290420293808}{357,020033121108} \times 100\%$$

$$= 1,04\%$$

The calcualtion indicates that in SID, Data Type INT is 1,04 faster than VARCHAR(8).

## 5. CONCLUSION

SID data, in this case alfanumeric data type, can be expressed as a string or integer. Seen from the standpoint of numbers, alfanumeric should be represented as a string. However, a practical approach shows that the speed and data storage space is more important than the definition of numbers. From the discussion on the comparison of storage space and access speed, the SID will be better if it is represented by an integer (INT) instead of a string (VARCHAR(8)). The representation with integer is 54% more efficient than VARCHAR (8) for space. In speed of access, the representation with integer is only 1,04% faster than VARCHAR (8). The integer type is still better than string type. Thus it can be concluded that SID with alphanumeric data type, the data representation using integer is better than that using string.

## ACKNOWLEDGEMENTS

The authors would like to thank Mr. Christanto Triwibisono, for the discusiion about the better performanace of Information System.

## Authors

Agus Pratondo received his bachelor degree in Informatics, Insitute of Technology Bandung (ITB), in 2001. He has several industrial experiences in some jobs position e.g. programming in C, PHP, and Manager of Software Quality Assurance. He joined the Department of Information System, Telkom Polytechnic in September 2007. He received master degree in Information Technology in 2008 from ITB. His research focused on information system, mobile application, and forecasting.

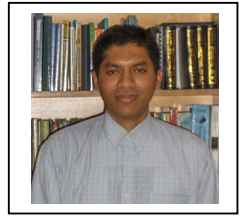